\documentclass[jgrga]{agutex}

\usepackage{graphicx}
\usepackage{amsmath}
\usepackage{dsfont}

\newcommand{\abs}[1]{\ensuremath{\left\lvert #1\right\rvert}}

\newcommand{\be}{\begin{equation}}
\newcommand{\ee}{\end{equation}}
\newcommand{\bs}{\begin{subequations}}
\newcommand{\es}{\end{subequations}}

\newcommand{\R}{\ensuremath{\mathds{R}}}

\newcommand{\pa}{\ensuremath{_\parallel}}
\newcommand{\se}{\ensuremath{_\perp}}

\newcommand{\De}{\ensuremath{\varDelta}}
\newcommand{\Om}{\ensuremath{\varOmega}}

\newcommand{\f}[1]{\ensuremath{\boldsymbol{#1}}}
\newcommand{\m}[1]{\ensuremath{\left\langle #1\right\rangle}}

\newcommand{\dd}[2][]{\ensuremath{\frac{\mathrm{d} #1}{\mathrm{d} #2}}}
\newcommand{\df}{\ensuremath{\mathrm{d}}}

\authorrunninghead{TAUTZ ET AL.}

\titlerunninghead{PARKER SPIRAL GEOMETRY}

\authoraddr{R. C. Tautz,
Zentrum f\"ur Physik und Astronomie, Technische Universit\"at Berlin, Hardenbergstra\ss e 36, D-10623 Berlin, Germany.
(rct@gmx.eu)}

\authoraddr{A. Shalchi, Department of Physics and Astronomy, University of Manitoba, Winnipeg, Manitoba R3T~2N2, Canada.
(andreasm4@yahoo.com)}

\authoraddr{A. Dosch, CSPAR, University of Alabama in Huntsville, Huntsville AL~35805, USA.
(ad@tp4.rub.de)}

\begin{document}

\title{Simulating Heliospheric and Solar Particle Diffusion\\
using the Parker Spiral Geometry}



\authors{R. C. Tautz, \altaffilmark{1} A. Shalchi, \altaffilmark{2} and A. Dosch \altaffilmark{3}}

\altaffiltext{1}{Zentrum f\"ur Physik und Astronomie, Technische Universit\"at Berlin, Hardenbergstra\ss e 36, D-10623 Berlin, Germany.}
\altaffiltext{2}{Department of Physics and Astronomy, University of Manitoba, Winnipeg, Manitoba R3T~2N2, Canada.}
\altaffiltext{3}{CSPAR, University of Alabama in Huntsville, Huntsville AL~35805, USA.}

\begin{abstract}
Cosmic Ray transport in curved background magnetic fields is investigated using numerical Monte-Carlo simulation techniques. Special emphasis is laid on the Solar system, where the curvature of the magnetic field can be described in terms of the Parker spiral. Using such geometries, parallel and perpendicular diffusion coefficients have to be re-defined using the arc length of the field lines as the parallel displacement and the distance between field lines as the perpendicular displacement. Furthermore, the turbulent magnetic field is incorporated using a WKB approach for the field strength. Using a test-particle simulation, the diffusion coefficients are then calculated by averaging over a large number of particles starting at the same radial distance from the Sun and over a large number of turbulence realizations, thus enabling one to infer the effects due to the curvature of the magnetic fields and associated drift motions.
\end{abstract}

\begin{article}

\section{Introduction}

It is accepted that, in general, the Solar magnetic field can be described through an Archimedean spiral as originally suggested by \citet{par58:spi}. Today, the Parker model is the starting point for countless theoretical---both analytical and numerical---investigations. Observations show general agreement with the model \citep{for96:uly}, especially in the ecliptic plane and regarding the most probable field direction, although the magnetic field lines are less tightly wound as predicted by the model. However, there are overall deviations from the Parker structure if the field is measured far away from the ecliptic plane, as has been done by the Ulysses spacecraft \citep{smi01:uly}. It has been shown that such can be explained by the motion executed by the foot points of magnetic field lines across coronal loop hole boundaries \citep{sch05:sub}, thus giving rise to a ``sub-Parker'' structure. Using numerical simulations, \citet{ril07:sim} confirmed the connection of a more radial magnetic field and coronal holes. Furthermore, at high Solar latitudes there are asymmetries in the azimuthal field component, and the field lines are generally more inclined towards the equator due to the interaction with Solar wind plasma \citep{for96:uly}.

The linear radial decrease of the (azimuthal) field component that was predicted by the Parker model was confirmed from the investigation of Pioneer~10 data \citep{par76:spi}, although some deviations have been known depending on the phase of the Solar cycle; for example, the radial component is independent on Solar latitude during Solar minima and maxima, and the overall field magnitude is smaller than expected \citep{smi04:mag}. Furthermore, near the Sun the magnetic field has to be described through a superposition of dipole, quadrupole, and current sheet structures, thus giving rise to extensive modifications of the Parker structure especially during Solar minima \citep{ban98:sol}. In defense of Parker, however, it should be noted that his model describes the field only outside of the zone where field lines execute a rigid-body rotation.

It has been argued that, due to some difficulties with vanishing magnetic field divergence, the standard Parker spiral has to be superposed by other field components, preferably homogeneous \citep{gol98:mag} with a strength of a fraction of nT. Such could be interpreted as the local galactic magnetic field, for which estimates are available through carefully distinguishing the excellently measured Solar field and the large-scale field as inferred from pulsar rotation rates \citep{ran89:gal}. However, it should be noted that the form of the magnetic field can be inferred from the study of suprathermal electrons (with energies larger than $70$\,eV), which are aligned with the heliospheric magnetic field (and have therefore been named ``strahl''), because such a particle distributions broadens with distance \citep{owe08:spi}. Independently from other models, such leads to a well-formed Parker geometry.

A modified model \citep{fis96:mag} includes the reversion of polarity and thus deviates from the Parker model because the magnetic field components now depend on the field polarity. From the solution of the transport equation, it was shown that the energy spectra of protons arriving at Earth show significant differences depending on the underlying magnetic field model. Similarly, the transport equation has been solved for solar energetic particle events (SEP) using both finite difference \citep{kal93:inh} and stochastic methods \citep{zha09:sol}. Such particles have sufficiently high energy to pose a threat to both astronauts and space infrastructure such as satellites. Based on a magnetic field in the form of a Parker spiral, the results underline again that perpendicular diffusion is important if one wants to understand the development of the SEP during the time it needs to arrive on Earth. Therefore, the knowledge of the magnetic field geometry is important if one attempts to understand the propagation of particles, which is also important for a prediction of ``space weather'' \citep{pom02:mag}.

The smooth, large-scale field described by the Parker---and similar, extended---models is, however, only half the truth; in reality, the heliospheric magnetic field is highly variable on all scales, which can be described in terms of waves, fluctuations, or more general turbulence models \citep{smi89:mag}. Thus, the spiral structure becomes evident only after (time and space) averaging of the magnetic field, showing that the turbulence strength is of the same order of magnitude as the background field. Moreover, as known from observations, the small- and middle-scale structures, can in turn influence the underlying spiral structure \citep{dal02:mid} by changing the inclination angle of the magnetic field lines. It is generally assumed that the dominant constituents are magnetohydrodynamic Alfv\'en waves \citep{smi89:mag,tau10:wav}, which propagate in opposite directions along the background magnetic field lines \citep{cab98:mhd}, and convective, two-dimensional structures \citep{sch95:alf}. However, other waves such as magnetosonic waves have also been used, thereby giving rise to anisotropies in the power spectra used to characterize the turbulence \citep{cha00:ani}. Furthermore, magnetosonic waves dominate the stochastic acceleration of charged particles \citep{sch98:ttd,tau10:wav}.

The motion of charged particles in such systems can be described by a diffusion process \citep[see][for detailed introductions]{rs:rays,sha09:nli}. Although the problem cannot be considered to be completely solved, significant progress has been made in describing the motion of energetic particles such as Cosmic Rays in a turbulent medium immersed in a homogeneous background magnetic field. For example, simulation results in different turbulence geometries \citep{gia99:sim,qin02a:rec,qin02b:sim,tau10:pad} could be reproduced using non-linear extensions \citep[e.\,g.,][]{mat03:nlg,sha04:wnl,sha06:enl,qin07:nle,tau08:soq,dos09:rel,sha10:uni} of the quasi-linear transport theory \citep{jok66:qlt}.

As soon as it comes to curved background fields, however, matters are even less understood. The concept of adiabatic focusing has attracted attention recently \citep[e.\,g.,][]{kun79:adf,spa81:adf,bie02:obs,sch08:adf}, which describes the modification of diffusion and transport equations due to a spatial gradient in the magnetic field. Other models simply prescribe a curved spatial geometry by accepting a background magnetic field with non-zero divergence \citep{min07:dri}. But a satisfying solution has not been achieved yet for real magnetic configurations that are not uniformly converging or diverging---such as the Parker spiral. In principle, the real field lines, which are distorted by the presence of turbulence, can be inferred using a field line wandering (or field line random walk) approach \citep{jok69:sto,mat95:tra,sha07:flr}. By specifying the turbulence power spectrum, a direct calculation of the stochastic field lines is then possible, showing excellent agreement with Helios~2 data. Furthermore, \citet{kob01:imf} showed that the diffusion tensor can be calculated analytically by applying two rotations to the parallel and perpendicular diffusion coefficients, although, in general, such a transformation will always be ambiguous.

Furthermore, the question of ``diffusivity'', i.\,e., whether the diffusion coefficients attain finite values, is still not solved satisfactorily. Whereas, in perpendicular scattering in magnetostatic slab turbulence (in Alfv\'enic slab turbulence, diffusion is recovered \citealt{sha07:alf,tau10:wav}), sub-diffusion is clearly established, it is generally assumed that, in isotropic turbulence, the particle motion is diffusive---although there are contradicting analytical results \citep{tau08:sem}. However, for anisotropic non-slab turbulence, such is not clear \citep{zim06:dif}. There are several cases where sub- or even super-diffusion has been found in numerical simulations \citep{tau10:sub} and in observations of electrons accelerated at interplanetary shocks \citep{per07:sup}. It has even been thought that the underlying process might not be a classic diffusion process but a L\'evy random walk \citep{zim05:ano}. However, such analyses would first require a stable description of the turbulence, because the results depend sensitively on the details of the turbulence model and its parameters.

In this article, the transport of test-particles is investigated using a numerical Monte Carlo technique \citep{gia99:sim,mic01:sim,tau10:pad}. The magnetic field is composed of a large-scale Parker-type field and a turbulent component, which is assumed to be isotropic. It is known from observations \citep{mat90:mal,bie94:pal,cha00:ani} that the turbulent magnetic field component is most likely not isotropic. There are even studies that incorporate such effects in sophisticated turbulence models \citep{sri94:tur,gol95:tur}. The problem, however, is that the basic transport theories are not sufficiently understood yet to allow for the isolation of small effects. Here, therefore, an isotropic turbulence spectrum will be used \citep{tau06:sta,tau08:soq} that uses a constant energy range. Numerically, the well-known method will be applied of superposing a homogeneous (in our case: the Parker spiral) magnetic field with a turbulent component calculated by the summation over plane waves with random directions of propagation and random phase angles \citep{tau10:pad}. Thereby, the diffusion coefficients are calculated from the mean square deviation. It will be shown how ``parallel'' and ``perpendicular'' particle displacements must be transformed using concepts from differential geometry. The numerical ansatz will be relativistically correct, although such might not be necessary for the energies considered. In a second paper \citep{tau10:whi}, the effects of Whistler wave turbulence will be studied, which is important for the formation of the ``strahl'' electrons that have been mentioned above. Here, in contrast, magnetostatic turbulence will be employed. Although, as has already been mentioned, plasma waves may have a severe impact on the diffusion coefficients especially if their electric fields are of significant magnitude, such a configuration represents the most basic case that has to be understood first before implementing more advanced features. The approach is, to some extent, comparable to a method used by \citet{pei06:spe}, although in their work focus was laid on the calculation of SEP onset times. It was found that, in the presence of turbulent field component, the onset times were generally reduced in comparison to a smooth Parker magnetic field.

The present paper is organized as follows: In Sec.~\ref{par}, the geometry of the Parker spiral is introduced and it is shown how appropriate modifications can be made to the (numerical) implementation of the simulation code that calculates the diffusion coefficients. In Sec.~\ref{sim}, the setup of the test-particle simulations and the turbulence generation are briefly explained. In Sec.~\ref{res}, the simulation results are presented and compared both to previous calculations and simulations in homogeneous background magnetic fields. Finally, Sec.~\ref{summ} provides a summary and conclusions regarding future work.

\section{Parker Spiral Geometry}\label{par}

According to \citet{par58:spi}, the large-scale spiral pattern of the Solar magnetic field in interplanetary space (without taking into account the turbulent component) is determined through an equation relating the radial distance from the Sun, $r$, and the azimuth angle in the ecliptic plane, $\phi$, as
\be\label{eq:par1}
\frac{r}{b}-1-\ln\!\left(\frac{r}{b}\right)=\frac{v_{\text{SW}}}{b\omega_\odot}\left(\phi-\psi\right),
\ee
where the parameter $\psi$ is the azimuth angle of the field line at the co-rotation radius. Furthermore, $v_{\text{SW}}\approx400\,$km/s is the outward velocity of the solar wind, $\omega_\odot$ is the angular velocity of the sun, and $b=46\times10^{-3}$\,AU is a distance beyond which any direct influence of the sun may be neglected \citep[cf.][]{par58:spi}. Physically, $b$ marks the end of the region, where the field lines execute a ``rigid rotation'' and, therefore, point radially outwards; i.\,.e., for $r<b$ the field lines point radially outward while they corotate with the Sun's surface and, thus, Eq.~\eqref{eq:par1} is not valid. Here, however, only the case $r>b$ will be considered.

\begin{figure}[t]
\centering
\includegraphics[width=80mm]{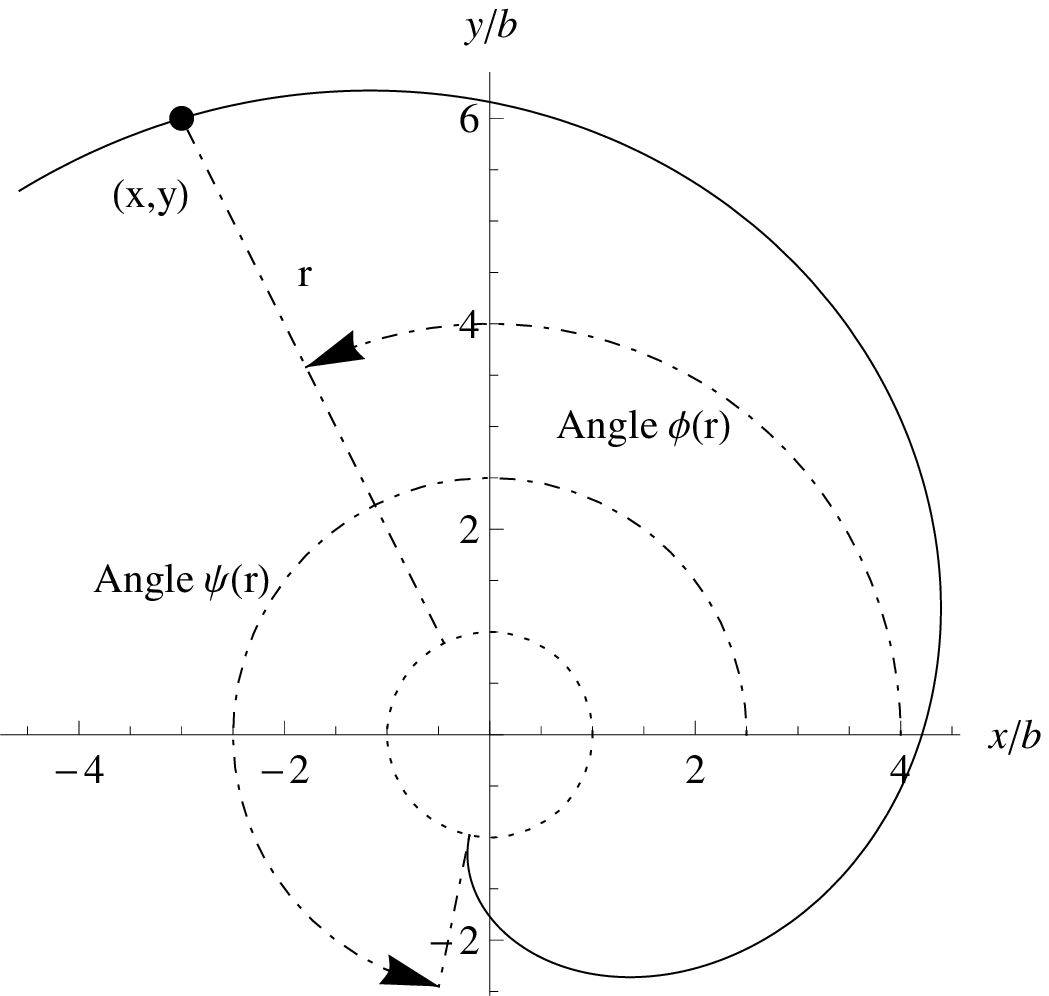}
\caption{Illustration of the angles used to describe the Parker spiral. The parameter $\psi$ corresponds to the angle at which the large-scale field line (solid line) starts, whereas the variable $\phi(r)$ determines the course of the field line in dependence on the radius, $r$.}
\label{ab:angles}
\end{figure}

\begin{figure}[t]
\centering
\includegraphics[width=80mm]{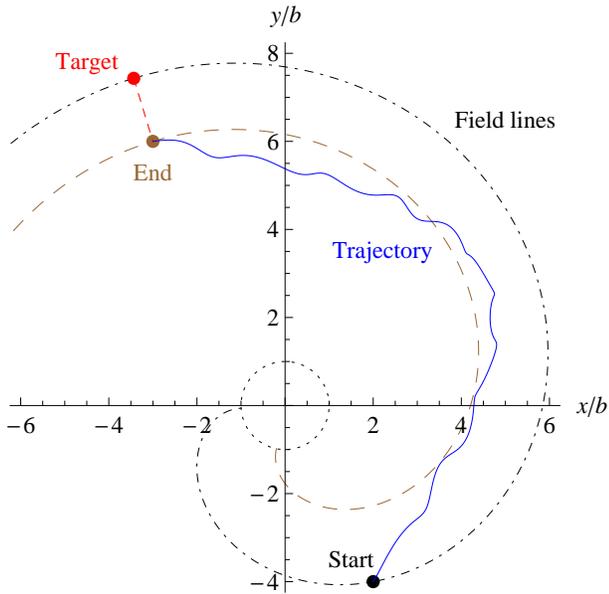}
\caption{(Color online) The meaning of ``parallel'' and ``perpendicular'' in Parker spiral geometry. Starting (black dot) on the black dash dotted field line, the particle trajectory (blue solid line) ends (brown dot) on the brown dashed field line. As an approximation, the perpendicular distance between the two field lines (red short-dashed line) is taken to be a straight line, where the target point (red dot) has to be calculated through minimization of the perpendicular distance.}
\label{ab:spiral}
\end{figure}

In what follows, quantities and relations will be referred to Figs.~\ref{ab:angles} and \ref{ab:spiral}, which illustrate the explanations given here. Solving Eq.~\eqref{eq:par1} for the azimuth angle $\phi$ as a function of the radius, $r$, yields
\be
\phi(r)=\psi+\frac{b}{\zeta}\left[\frac{r}{b}-1-\ln\!\left(\frac{r}{b}\right)\right],
\ee
with $\zeta=v_{\text{SW}}/\omega_\odot\approx1$\,AU \citep{zan04:dif}.

Let us now consider an arbitrary particle in the ecliptic plane (black dot) at the cartesian coordinates ($x_{\text s},y_{\text s}$) denoting the starting point on a magnetic field line (black dash dotted line). The starting point may be expressed through the new coordinates $(r_{\text s},\phi_{\text s})$ using the transformations
\bs
\begin{align}
r_{\text s}&=\sqrt{x_{\text s}^2+y_{\text s}^2}\label{eq:rstart}\\
\phi_{\text s}&=\arctan\left(\frac{y_{\text s}}{x_{\text s}}\right),
\end{align}
\es
where, on evaluating the $\arctan$ function, attention has to be paid to choosing the right quadrant. The inner field line azimuth angle, $\psi$, can be determined by back-tracing the field line to the radius $r=b$. Since the starting point lies on that field line, $\psi$ can simply be calculated by
\be
\psi=\phi_{\text s}-\frac{b}{\zeta}\left[\frac{r_{\text s}}{b}-1-\ln\!\left(\frac{r_{\text s}}{b}\right)\right],
\ee
which is visualized in Fig.~\ref{ab:angles}.

Usually the end point of the particle trajectory, $(x_{\text e},y_{\text e}$), will be located on a different (but nearby) field line (see brown dot on the brown dashed line in Fig.~\ref{ab:spiral}). Particularly, the particle has drifted from one field line to another due to diffusion processes. According to the definition of the parallel and perpendicular (running) diffusion coefficients,
\be
\kappa_{\parallel,\perp}(t)=\frac{\m{(\De s_{\parallel,\perp})^2}}{2t},
\ee
one has to calculate the parallel and perpendicular mean square displacements, respectively, which are denoted by $\langle\De^2\rangle$. The perpendicular displacement is associated with the shortest distance---perpendicular to the field lines---between the ending point and the initial field line, leading to the target point (red dashed line and red dot). In a uniform magnetic field such is simply a straight line perpendicular to the background magnetic field lines. Likewise, the parallel displacement is the shortest distance between the starting point and the target point; in a uniform magnetic field, it would be a straight line, too. But how does that have to be translated for a curved background magnetic field such as that described by the Parker spiral?

The natural answer is motivated by differential geometry. The perpendicular displacement is characterized by a curve perpendicular to all field lines, which runs through the end point and the target point. As a first approach it is assumed that such a curve can be approximated by a straight line as in Euclidean geometry (red dashed line), based on the fact that the turbulence in interplanetary space is moderate \citep{bie94:pal} and, therefore, perpendicular diffusion is weak, too. Furthermore, it was shown by \citet{sha09:ani} that, even in strong isotropic turbulence, one generally has $\kappa\se\ll\kappa\pa$.

\begin{figure}[t]
\centering
\includegraphics[width=80mm]{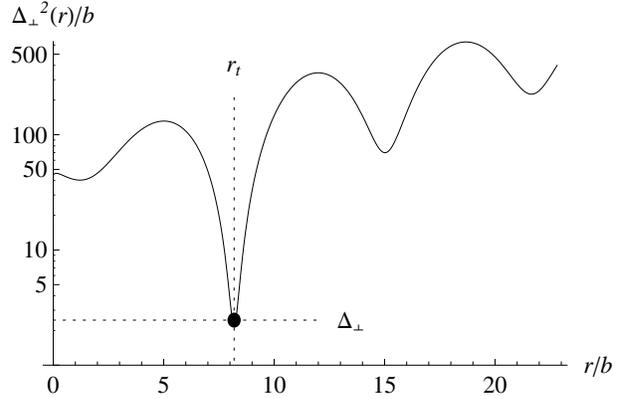}
\caption{The function from Eq.~\eqref{eq:rmin}, the minimum of which (black dot) has to be found in order to calculate the radius of the target point, $r_{\text t}$, and the associated perpendicular and parallel displacements, $\De s\se$ and $\De s\pa$, respectively. The correct minimum is that next to the starting point, for which $r_{\text s}/b=4.47$.}
\label{ab:minimum}
\end{figure}

This also motivates the assumption that the particle trajectory ends on nearby field lines and states that the parallel displacement is given through the arc length of the initial (black dash dotted line) field line to a (red dot) target point, where the shortest distance to the (brown dot) particle end point is perpendicular on all field lines. Hence, instead of a straight line the ``shortest distance'' is now a geodesic following the geometry of the Parker spiral. As an approximation, however---owed to the fact that the turbulence in interplanetary space is weak and that, therefore, perpendicular diffusion is weak, too---the calculation of the target point proceeds as follows: One simply calculates the point from which the distance---this time taken as a straight line as in Euclidean geometry---to the particle trajectory's end point is minimal (red dashed line). Then, the perpendicular displacement, denoted by $\De s\se$, is simply given through the distance between the target point, $r_{\text t}$, and the end point, $r_{\text e}$, as
\be\label{eq:rmin}
\De s\se^2=\min_{r_{\text t}\in\R^+}\Bigl\{\left[r_{\text t}\cos\bigl(\phi(r_{\text t})\bigr)-x_{\text e}\right]^2+\left[r_{\text t}\sin\bigl(\phi(r_{\text t})\bigr)-y_{\text e}\right]^2\Bigr\},
\ee
which function is visualized in Fig.~\ref{ab:minimum}.

The only intricacy is to find the correct minimum, since the function in Eq.~\eqref{eq:rmin} has multiple local minima for $\phi+2\pi k$ with $k$ approximately an integer number. Hence, a method in three steps will be used. First step: to identify a minimum of any function $f$, three points $a<b<c$ are needed with $f(b)<f(a)$ and $f(b)<f(c)$ are necessary---which is called ``bracketing'' of the minimum \citep[p.~490]{pre07:nr3}. Assuming $\phi_{\text t}\approx\phi_{\text e}$ and assuming that the particle will not complete one or more full orbits around the Sun, one can therefore search for a minimum starting from $r_{\text s}$. By iteratively solving for $r(\phi_{\text e}\approx\phi_{\text t})$ starting with $r_0=r_{\text s}$ one has, therefore,
\be
r_{i+1}=b\left[1+\ln\!\left(\frac{r_i}{b}\right)\right]+\zeta\left(\phi_{\text e}-\psi\right),
\ee
which will be stopped after a few steps. For very small displacements $(r_{\text e},\phi_{\text e})\approx(r_{\text s},\phi_{\text s})$, i.\,e., after short times, however, the iteration can yield wrong results. Second step: In that case, a fail-safe method will determine an approximate bracketing condition from calculating $r(\phi_{\text s}-2\pi)\leq r_{\text s}\leq r(\phi_{\text s}+2\pi)$, where again use is made of the assumption that a given particle will not complete a full orbit around the Sun. Third step: Based on the bracketing of the minimum, it is then a simple task to calculate the exact minimum using, e.\,g., Brent's method \citep[pp.~496--502]{pre07:nr3} without derivative information [see also \citealt{for77:mat}, \S8.2, \citealt{bre02:min}, Chpt.~5].

Once the target point has been determined, the parallel displacement, $\De s\pa$, is calculated through the arc length of the field line between the target point and the starting point, yielding
\be\label{eq:de_pa}
\De s\pa=\mathcal L(r_{\text t})-\mathcal L(r_{\text s}),
\ee
where the arc length function, $\mathcal L$, can be calculated analytically and is given through
\bs
\begin{align}
\mathcal L(r)&=\int^r\df s\,\sqrt{1+s^2\left(\dd[\phi]s\right)^2}\\
&=\frac{r-b}{2}\,\frac{\sqrt{(r-b)^2+\zeta^2}}{\zeta}+\frac{\zeta}{2}\;\text{Arsinh}\left(\frac{r-b}{\zeta}\right).
\end{align}
\es
For instance, the trajectory parameters from Fig.~\ref{ab:spiral} (with $\zeta=1$)have been used to obtain $\De s\pa/b=20.2$ [by using Eq.~\eqref{eq:de_pa}] and $\De s\se/b=2.2$.

Until here, only the ecliptic plane---which, in the simulation box, corresponds to the $x$-$y$ plane---has been considered. To incorporate drift motion, the vertical displacement, $\De z=z_{\text e}-z_{\text s}$, is calculated so that, from the comparison of $\De s\se$ and $\De_z$, the drift effect can be inferred.

Furthermore, the (background) magnetic field components, which are designed to cover mainly the ecliptic plane, are described using spherical coordinates through \citep[see][]{par58:spi,bur04:mag}
\bs\label{eq:par_magn}
\begin{align}
B_r(r,\theta,\phi)&=B_0\left(\frac{b}{r}\right)^2\\
B_\theta(r,\theta,\phi)&=0\\
B_\phi(r,\theta,\phi)&=B_0\left(\frac{b}{r}\right)^2\frac{r-b}{\zeta}\,\sin\theta,
\end{align}
\es
with $B_0=1830$\,nT \citep{zan04:dif}. Note that the strength of the magnetic field is maximal in the ecliptic plane, and that no care has been taken of the current sheet and the polarity reversal \citep[cf.][]{fis96:mag,bur04:mag}. For small $r$, the field is purely radial and scales with $r^{-2}$, whereas, according to Eqs.~\eqref{eq:par_magn}, for large $r$ the field becomes more and more azimuthal and scales with $r^{-1}$, in agreement with Pioneer~10 data \citep{par76:spi}; see also Fig.~\ref{ab:spiral} and Fig.~6 of \citet{par58:spi}.

When the diffusion coefficient is calculated in the way described here, the curved space defined by the large-scale magnetic field is mapped to a Euclidean space. The diffusion coefficient can include the additional terms coming out of the divergence in the curved space. Thus, the diffusion coefficient carries a meaning different from local diffusion coefficient defined in the curved space. Note also that, until here, only the smooth spiral field lines have been considered. Both in the real world and in the simulation code, the background field will be superposed by a turbulent component. However, diffusion coefficients will be calculated in reference to the mean field component.

\section{Monte-Carlo Simulations}\label{sim}

For the simulation of cosmic ray scattering processes, a modified version of the recently developed \textsc{Padian} code \citep{tau10:pad} has been used, which traces the trajectories of a large number of test particles (typically $10^4$) for a sufficiently long time (typically $10^4$ Larmor orbits). In doing so, it can be decided whether the transport is diffusive, unperturbed, or subdiffusive, and the diffusion coefficients can be determined. For the integration of the equation of motion, an integration algorithm with adaptive step sizes such as the Bulirsch-Stoer method \citep[see][pp.~921--928]{sto02:num,pre07:nr3} is used, which limits the relative deviation in the particle rigidity, $\De R/R$, to be smaller than $10^{-3}\%$. Here, the particle is expressed as momentum per unit charge times speed of light, i.\,e., $\f R=\f pc/e$. Note the difference to several previous articles \citep[e.\,g.,][]{tau08:soq,tau10:pad}, where a dimensionless ``rigidity''-like quantity was defined through $\f R=\gamma m\f v/(\Om\ell_0)$, where $\Om$ denotes the gyro-frequency.

In comparison to the original formulation \citep{tau10:pad}, the equation of motion---i.\,e., the Newton-Lorentz equation---has to be slightly modified due to the non-constant background magnetic field. Furthermore, SI units are used so that the rigidity and the magnetic field strengths can be given in megaVolt (MV) and nanoTesla (nT), respectively. The trajectory $\f x(t)$ (normalized to a characteristic length scale $\ell_0$; see below) is therefore calculated as a function of the dimensionless time $\tau=vt/\ell_0$ using
\bs\label{eq:NL}
\begin{align}
\dd\tau\,\frac{\f x}{\ell_0}&=\frac{1}{R}\,\f R\\
\dd\tau\,\f R&=\frac{a}{R}\,\f R\times\left[B_0(r)\,\hat{\f e}_{B_0}+\delta B(r)\,\hat{\f e}_{\hspace{0.2pt}\delta\hspace{-0.2pt}B}\right], \label{eq:NLb}
\end{align}
\es
where Eq.~\eqref{eq:NLb} corresponds to $\dot{\f p}=q\f v\times\f B$ (SI units). The scaling factor is given by $a=\ell_0c/10^{15}\approx1345.45\,\text m^2/\text s$ for typical parameters (see below). Note that Eqs.~\eqref{eq:NL} are valid for all particle species and that the rest frame of the Sun has been used.

The unit vectors $\hat{\f e}_{B_0}$ and $\hat{\f e}_{\hspace{0.2pt}\delta\hspace{-0.2pt}B}$ point in the directions of the Parker field and the turbulent magnetic field, respectively. In the ecliptic plane, the total background magnetic field strength is, approximately, determined through Eq.~\eqref{eq:par_magn} as
\be\label{eq:B0}
B_0(r)=1830\,\text{nT}\left(\frac{b}{r}\right)^2\sqrt{1+\left(\frac{r}{\zeta}\right)^2}.
\ee
Following \citet[Sec.~3.1]{zan96:tur}, a WKB approximation \citep[see also][]{mat94:wkb} leads to an equation for the turbulent magnetic field energy per mass, $E_{\delta B}=\delta B^2/(8\pi\rho)$, as
\be\label{eq:EdB}
\dd[E_{\delta B}]r+\frac{E_{\delta B}}{r}=0,
\ee
yielding $E_{\delta B}\propto r^{-1}$. Equation~\eqref{eq:EdB} was motivated by the Wal\'en relation \citep{mat02:wal}, i.\,e., $\De\f B\propto\rho\De\f v$, where $\De\f B$ and $\De\f v$ are changes in the magnetic field and the plasma velocity, respectively. Based on the Wal\'en relation, the ratio of kinetic and magnetic energy (both per mass) are constant. Therefore, with the additional assumption that, in the ecliptic plane, the mass density behaves as $\rho\propto r^{-2}$, one has
\be
\left(\frac{\delta B}{\delta B_{\text{ref}}}\right)^2=\left(\frac{r_{\text{ref}}}{r}\right)^3,
\ee
where $\delta B_{\text{ref}}=4$\,nT is the turbulent field strength at $r_{\text{ref}}=1$\,AU. Accordingly, the turbulent magnetic field strength is given through \citep[Appendix~B]{zan04:dif,sha10:ana}
\be
\delta B(r)=4\,\text{nT}\left(r\,[\text{AU}]\right)^{-3/2},
\ee
where $r$ is the radial distance from the Sun in astronomical units (AU). Based on Eq.~\eqref{eq:B0}, the relative strength of turbulent and background magnetic field depends on $r$ as
\be
\frac{\delta B}{B_0}\approx\sqrt{\frac{r\,[\text{AU}]}{1+\left(r\,[\text{AU}]\right)^2}}.
\ee

The turbulence model, which, to some extent, is based on a model used by \citet{gia99:sim}, generates random magnetic fluctuations by the superposition of a large number $N$ of plane waves (typically $512$) with random directions of propagation, random phases, and amplitudes as prescribed by the turbulence spectrum of \citet{tau06:sta,tau08:soq} that has the form
\be\label{eq:spect}
G(k)=\left[1+\left(\ell_0k\right)^2\right]^{-\nu}\approx\left(\ell_0k\right)^{-5/3},
\ee
where the parameter $\nu=5/6$ denotes the Kolmogorov spectral index for a power-law $k^{-5/3}$ \citep[cf.][]{pod07:spe} and where $\ell_0$ is the isotropic turbulence bend-over scale. Note that, in the Solar system, the bend-over scale, $\ell_0$, may depend on the position \citep{bru05:sol}. The spectrum from Eq.~\eqref{eq:spect} has a constant energy range (i.\,e., the range $k<\ell_0^{-1}$). For a more general spectrum, a formula similar to that of Eq.~\eqref{eq:spect} can be found, e.\,g., in \citet{sha09:flr}.

The spatially fluctuating but time-independent (i.\,e., magnetostatic) magnetic field is then calculated as
\be\label{eq:dB}
\delta\f B(x,y,z)=\text{Re}\sum_{n=1}^N\f e'\se A(k_n)e^{i\left(k_nz'+\beta_n\right)},
\ee
where the sum extends over $N$ logarithmically spaced wavenumber values $k_n$. The amplitude function $A(k_n)\propto\sqrt{G(k_n)}$ is related to the turbulence spectrum and $\f e'\se$ is a unit vector in the direction perpendicular to $z'$. The primed coordinates are obtained from a rotation matrix, whose angles are randomly generated for each summand $n$. The parameter $\beta$ is the random phase for each wave mode. Thus, Eq.~\eqref{eq:dB} generates an isotropic, time-independent (i.\,e., magnetostatic) turbulent magnetic field. Accordingly, the orientation of $(x,y,z)$ is irrelevant; however, for convenience the $z$ direction is kept perpendicular to the ecliptic plane. The divergence of $\delta\f B$ is kept zero due to the fact that $\f e'\se\perp\f k_n$ for all summands, which corresponds to $\nabla\cdot\delta\f B=0$ in Fourier space.

Care must be taken about the minimum and maximum wave numbers. Whereas the analytical form of the spectrum from Eq.~\eqref{eq:spect} extends over all wavenumbers, such is not possible in computer simulations. There are two major conditions that have to be fulfilled, which are: (i) the resonance condition stating that there must exist a wavenumber so that $k\mu R_{\text L}=1$ is fulfilled; and (ii) the time scale condition stating that $\Om t_{\text{max}}k_{\text{min}}R_{\text L}<1$. The second condition requires a maximum turbulence scale (defined through $k_{\text{min}}^{-1}$) to be larger than the distance traveled by the particle to ensure the particle cannot move out of the system.

From simulations in a homogeneous background magnetic field \citep[e.\,g.,][]{tau09:per}, one knows that both conditions~(i) and (ii) do not depend on the turbulence bend-over scale $\ell_0$; instead, they depend on the maximum and minimum scale of the system (given by the minimum and maximum wavenumber, respectively). Condition~(ii) therefore states that the particles must not travel farther than the system scale $L_{\text{max}}=k_{\text{min}}^{-1}$. Although, in the simulation code, the turbulence is generated wherever the particle position is, particles start to free-stream once the condition is violated. In that case, one finds $\kappa\se\propto t^{-1}$, which is equivalent to a constant perpendicular mean square displacement and indicates a purely parallel motion.

\begin{figure}[t]
\centering
\includegraphics[width=80mm]{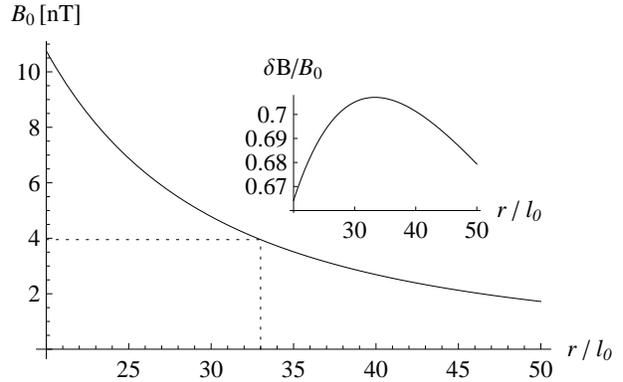}
\caption{The decrease of the background magnetic field strength with radial distance from the Sun as prescribed by the Parker spiral through Eqs.~\eqref{eq:par_magn}. At $r=r_{\text s}=33\,\ell_0$, one has $B_0=1.68$\,nT. The inset shows the relative strength of the turbulence in terms of the background magnetic field.}
\label{ab:magn}
\end{figure}

In Fig.~\ref{ab:magn}, the magnetic field strength is shown as a function of the distance from the Sun. In the \textsc{Padian} simulation code, the field strength is usually normalized to unity; here, however, the field strength decreases with increasing distance. Furthermore, the (average) magnitude of the turbulent field is shown in comparison to the background field strength, indicating a maximum at 1\,AU.

By integrating the particle trajectory as given by Eq.~\eqref{eq:NL}, the mean square displacement is then determined as described in Sec.~\ref{par}. Finally, the diffusion coefficients and the mean free paths are calculated by averaging over all particles and all different turbulence realizations.

\section{Simulation Results and Comparison}\label{res}

For the simulation runs, the particle energies were chosen in the range $10^1\,\text{MV}\leq R\leq10^4$\,MV. All runs were carried out using a total number of 1000 particles in 25 different turbulence realizations. It is interesting that, in contrast to simulations involving homogeneous background magnetic fields, here small particle energies require considerably longer computation times than high energies.

\begin{figure}[t]
\centering
\includegraphics[bb=190 262 385 565,width=70mm]{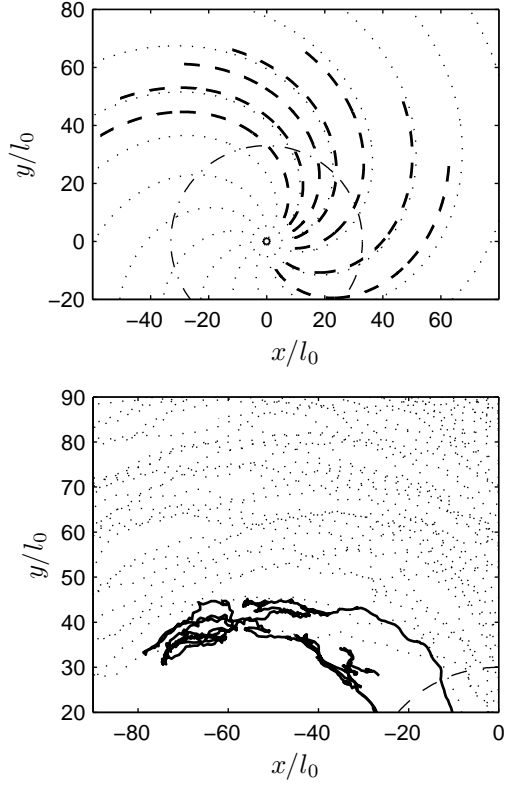}
\caption{Particle trajectories without (upper panel, dashed lines) and with (lower panel, solid lines) magnetic turbulence. As a guide to the eye, the magnetic field lines according to the Parker spiral (with, in the lower panel, superposed by the turbulent component) are shown as dotted lines. Furthermore, the dash dotted line roughly marks the Earth orbit around the Sun.}
\label{ab:traj}
\end{figure}

Furthermore, all particles are assumed to start in the ecliptic plane with an equal initial radial distance to the sun of $r_{\text s}=10\ell_0$. Since the bend-over scale is usually taken to be $\ell_0=0.03$\,AU, a heliocentric distance of 1\,AU corresponds to $33\ell_0$. While, in the absence of a turbulent magnetic field component, the particles follow the magnetic field lines, there is extensive scattering if the magnetic field is turbulent (see Fig.~\ref{ab:traj}). Especially at low particle energies, there even seems to be confinement-like mechanisms (not shown in the figure), where particles repeatedly move back and forth along the same magnetic field line without covering a large net distance.

\begin{figure}[t]
\centering
\includegraphics[width=85mm]{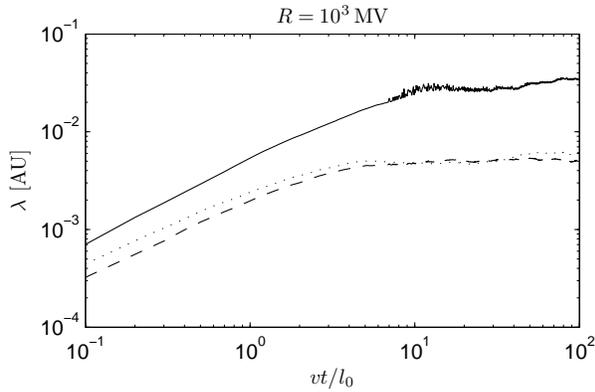}
\caption{The mean free paths normalized to the astronomical unit, $\kappa/(v\ell_0)$, as a function of the dimensionless simulation time, $vt/\ell_0$. The solid line shows the parallel mean free path (i.\,e., the coefficient of the diffusion along the curved magnetic field lines of the Parker spiral), whereas the dashed line show the mean free path perpendicular but in the ecliptic plane. The dotted line shows the vertical mean free path normal to the ecliptic plane, which is defined in the classic way, i.\,e., $\lambda_z=3\langle(\De z)^2\rangle/(2vt)$.}
\label{ab:Tlambda}
\end{figure}

In Fig.~\ref{ab:Tlambda}, the resulting mean free paths in the directions parallel, perpendicular, and vertical to the Sun's magnetic field are shown as a function of the normalized dimensionless time, $vt/\ell_0$. (Note that, here, the term ``perpendicular'' is used to describe a direction in the ecliptic plane, whereas ``vertical'' is oriented normal to the ecliptic plane.)

The comparison of the perpendicular and vertical mean free paths shows no qualitative difference, which leads to the conclusion that a drift effect can be neglected. The result that turbulent magnetic fields suppress particle drift motions is supported by the work of \citet{min07:dri}, who found that, if scattering processes are important, all drift motions are superseded by diffusion.

Furthermore, note that all three mean free paths in Fig.~\ref{ab:Tlambda} seem to approach a constant, thus indicating a diffusive behavior. In contrast to a configuration with a homogeneous background magnetic field as investigated in \citet{tau10:sub}, there is no indication here of a (even slightly) subdiffusive behavior---at least not for sufficiently high rigidities $R>10$\,MV.

In Fig.~\ref{ab:Rlambda}, the parallel mean free path is shown as a function of the particle rigidity. Furthermore, a comparison is shown to the Palmer consensus range (shaded box) consisting of mean free path values from various observational studies \citep[see][and references therein]{pal82:con,bie94:pal}.

Note that all simulation results were obtained for an initial radial distance $r_{\text s}=10\ell_0$. However, the difference to additional simulation runs with $r_{\text s}=30\ell_0$ turned out to be negligible. Therefore, the radial dependence of the magnetic field strength seems to have only a small effect on the mean free paths---in contrast to the turbulent magnetic field. According to quasi-linear theory \citep[e.\,g.,][]{jok66:qlt}, the turbulence strength relates to the parallel mean free path as $B_0/\abs{\delta B}\propto\sqrt{\lambda\pa}$, which behavior has also been observed in simulations \citep{tau10:wav} for intermediate turbulence strengths; for strong turbulence, such may be different \citep[cf.][]{sha09:ana}.

In Fig.~\ref{ab:ratio}, the ratio of the perpendicular/vertical and parallel mean free path is shown as a function of the particle rigidity. In agreement with previous work \citep[e.\,g.,][]{gia99:sim,sha04:wnl,zan04:dif,tau06:sta}, $\lambda\se/\lambda\pa$ tends to be slightly reduced as the particle rigidity increases. The overall values are somewhat higher than normally observed, because, using the estimate of \citet{bie04:nlg}, i.\,e., $\lambda\se/\lambda\pa\approx(\delta B/B_0)^2/6$, one has, at a radial distance of 10 and 20\,AU yields values for $\lambda/\se/\lambda\pa$ of approximately 0.05 and 0.07, respectively, whereas the simulation results are all well above $\lambda\se/\lambda\pa>0.1$ (see Fig.~\ref{ab:ratio}).

Finally, the ratio of vertical and perpendicular mean free paths is shown in Fig.~\ref{ab:drift}, illustrating that, although there is are noticeable uncertainties, the ratio is in agreement with unity. Therefore, the drift effects can be considered as negligible as already pointed out by \citet{min07:dri}.

\section{Summary and Conclusion}\label{summ}

In this article, particle scattering and diffusion parameters have been calculated for the Solar system. The investigation was based on the Parker spiral model for the background magnetic field together with a WKB model to account for the turbulent magnetic field component, which varies with the radial distance. For the first time, the Parker geometry of the Sun's magnetic field has been implemented in a test-particle simulation code.

\begin{figure}[t]
\centering
\includegraphics[width=85mm]{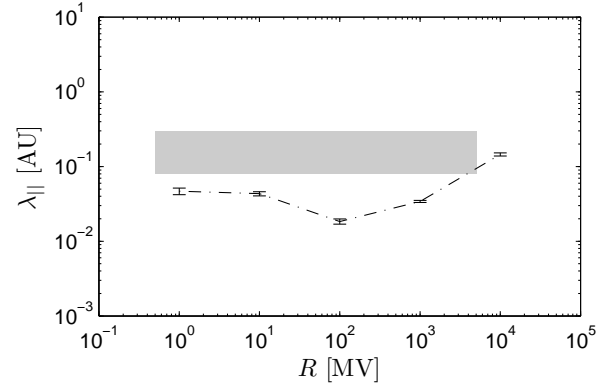}
\caption{The mean free path as a function of the particle rigidity. The error bars indicate the simulation results, and the shaded area illustrates the \citet[see also \protect\citealt{bie94:pal}]{pal82:con} consensus range.}
\label{ab:Rlambda}
\end{figure}

\begin{figure}[t]
\centering
\includegraphics[width=85mm]{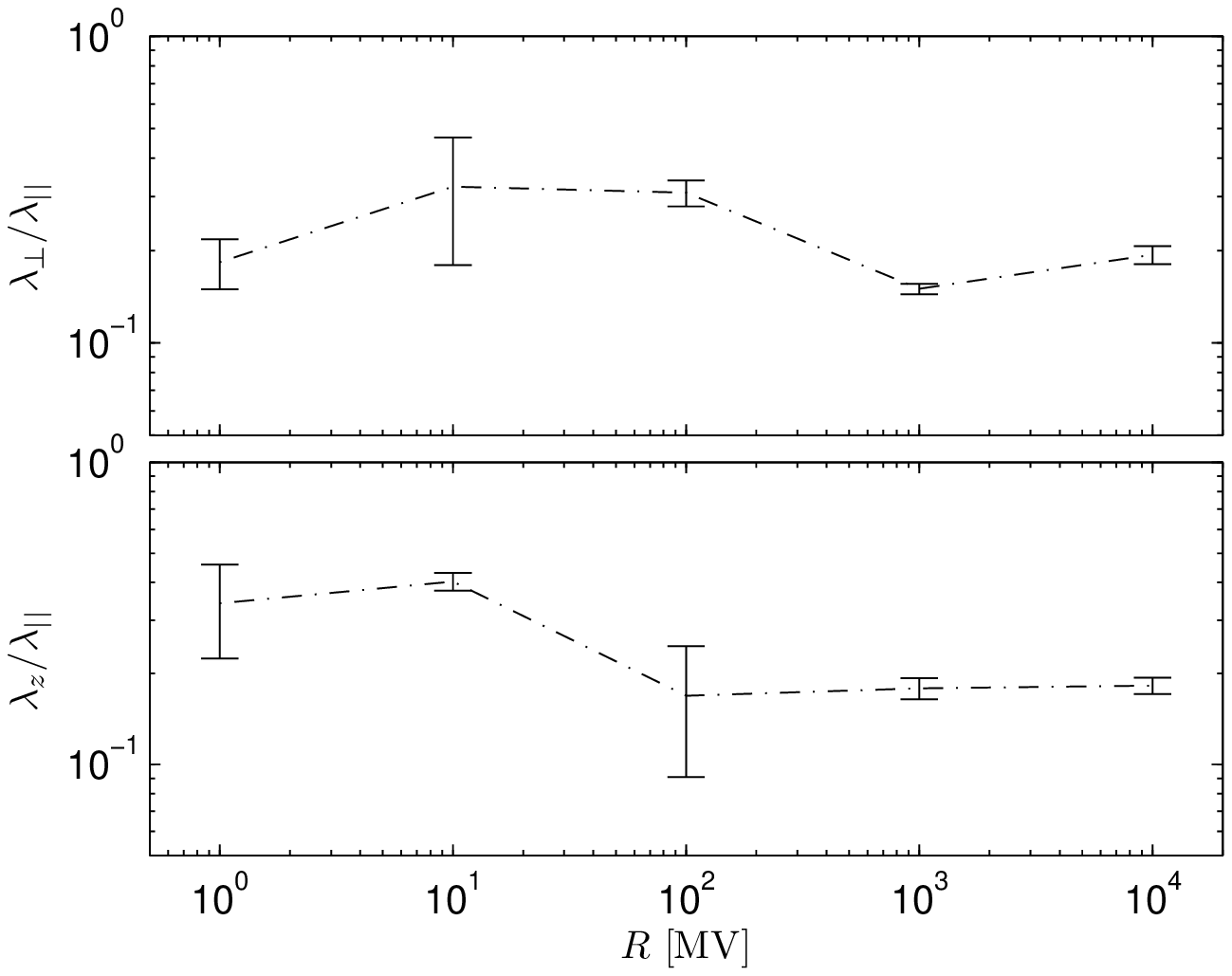}
\caption{The ratio of the perpendicular and the parallel mean free path (upper panel) and the ratio of the vertical and the parallel mean free path (lower panel). The error bars are calculated from the individual estimated mean errors of the individual mean free path components.}
\label{ab:ratio}
\end{figure}

\begin{figure}[t]
\centering
\includegraphics[width=85mm]{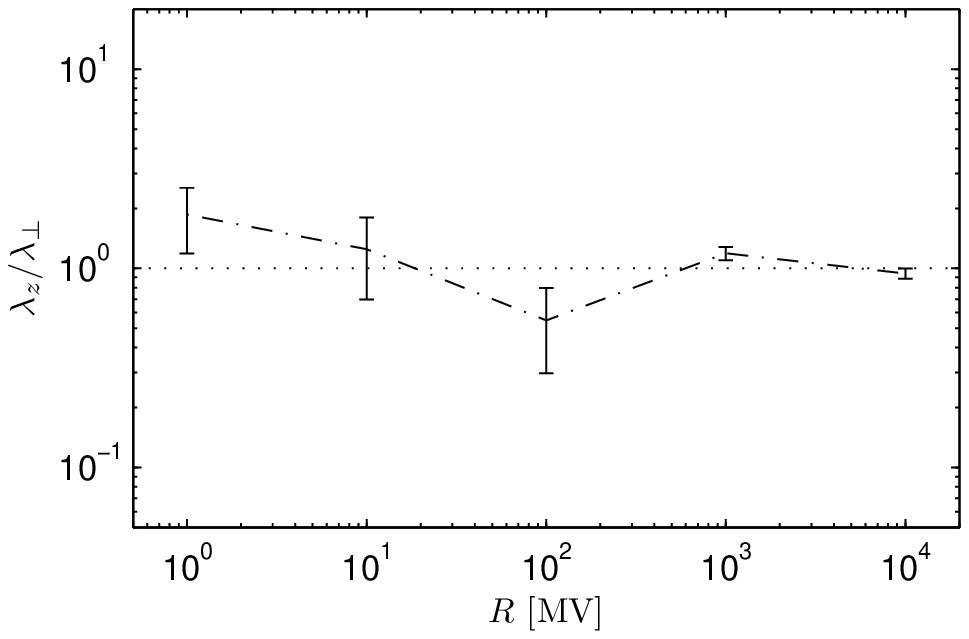}
\caption{The ratio of the perpendicular and the parallel mean free path (upper panel) and the ratio of the vertical and the parallel mean free path (lower panel). The error bars are calculated from the individual estimated mean errors of the individual mean free path components.}
\label{ab:drift}
\end{figure}

The prediction of negligible drift motions could be confirmed (Fig.~\ref{ab:drift}). It is obvious from Fig.~\ref{ab:Rlambda}, the results for the parallel mean free path shown here are too small, except for high rigidities $R\geq10^4$\,MV. There are three possible explanations:

First, in the present version an isotropic turbulence model has been used. Usually, a slab/2D model consisting of two turbulence components with wavevectors parallel and perpendicular to the background magnetic field are used for applications in the Solar system \citep{bie94:pal,bie96:two}, which increases the parallel mean free path \citep[e.\,g.,][]{tau10:sub}. On the other hand, it has been shown by \citet{wei10:mal} that a more realistic turbulence model, which overcomes the singularities inherent in the slab/2D formulation, can reproduce data calculated from spacecraft measurements better. However, the turbulence model that was designed in the style of the so-called Maltese cross \citep{mat90:mal} has not yet been implemented into simulation codes.

Second, the WKB model that has been used for the strength of the turbulent magnetic field could overestimate the true turbulence level. To allow for better adjustment of the turbulence parameters, more measurements of the turbulence level at different radial positions would be necessary.

Third, for other choices of the correlation scales, especially for different scales in the directions parallel and perpendicular to the background magnetic field, the resulting transport parameters would be different \citep[see also][]{pei10:hel}. Especially, a different choice of parallel and perpendicular turbulence bend-over scales could modify the resulting mean free paths \citep[Sec.~5.4.4]{mat03:nlg,sha09:nli}.

Future work should, therefore, try to implement a Maltese cross based turbulence model to account for the various anisotropies between the parallel and the perpendicular turbulent magnetic field components \citep[e.\,g.,][]{nar10:wav,che10:ani}. The effects due to the Solar wind would also be worth studying. Furthermore, a reliable description is necessary of the turbulent magnetic field strength on the radial distance from the Sun.

\begin{acknowledgments}
The work of R.\,C. Tautz was partially supported by the German Academy of Natural Scientists Leopoldina Fellowship Programme through grant LDPS 2009-14 funded by the Federal Ministry of Education and Research (BMBF).
\end{acknowledgments}


\newpage

\end{article}
\end{document}